\documentclass[%
 reprint,
 amsmath,amssymb,
 aps,
]{revtex4-2}

\usepackage{graphicx}% Include figure files
\usepackage{dcolumn}% Align table columns on decimal point
\usepackage{bm}% bold math
\usepackage[utf8x]{inputenc}
\usepackage{float}
\usepackage{color}
\usepackage{booktabs}
\usepackage{multirow}
\usepackage{diagbox}
\begin{document}

%\preprint{APS/123-QED}

\title{Nonrelativistic Treatment of Fully-Heavy Tetraquarks as Diquark-Antidiquark States}% Force line breaks with \\

\author{Halil Mutuk}
 \email{hmutuk@omu.edu.tr}
 \affiliation{ Department of Physics, Faculty of Science and Letters, Ondokuz Mayis University, 55139, Samsun, Turkey}


\begin{abstract}
The goal of the present work is to obtain a reliable estimate of the masses of the ground and radially excited states of fully-heavy tetraquark systems. In order to do this, we use a nonrelativistic model of tetraquarks which are assumed to be compact and  consist of diquark-antidiquark pairs. This nonrelativistic model is composed of Hulthen potential, a linear confining potential and spin-spin interaction. We computed ground, first, and second radially excited $cc\bar{c}\bar{c}$ and $bb\bar{b}\bar{b}$ tetraquark masses. It was found that predicted masses of ground states of $cc\bar{c}\bar{c}$ and $bb\bar{b}\bar{b}$ tetraquarks are significantly higher than the thresholds of the fall-apart decays to the lowest allowed two-meson states. These states should be broad and are thus difficult to observe experimentally. First radially excited states are considerably lower than their corresponding (2S-2S) two-meson thresholds. We hope that our study may be helpful to the experimental search for ground and excited $cc\bar{c}\bar{c}$ and $bb\bar{b}\bar{b}$ tetraquark states.

\end{abstract}

%\keywords{Tetraquark, Potential Model}%Use showkeys class option if keyword
\maketitle

%\tableofcontents

\section{\label{sec:level1}Introduction}

Quark model describes ordinary mesons as ($q\bar{q}$) systems and baryons as ($qqq$) systems in terms of quarks $q$ and antiquarks $\bar{q}$. In addition to quark model, the existence of multiquark states such as tetraquarks ($q\bar{q}q\bar{q}$ or $qq\bar{q}\bar{q}$), pentaquarks ($q\bar{q}qqq$), and structures with more  quarks was proposed decades ago \cite{GellMann:1964nj,Jaffe:1976ig,Jaffe:1976ih,Ader:1981db}. These new structures present quantum numbers, masses, flavours, decay channels and widths in the experiments which cannot be fitted into conventional quark model. Now they are called exotic hadrons. Concerning tetraquarks, the first discovery of exotic states was made in 2003 by Belle Collaboration in the charmonium sector \cite{Choi:2003ue}. This state was named as $X(3872)$ (now referred to name as $\chi_{c1}(3872)$) with a mass of ($3872.0 \pm 0.6$ MeV). Its extraordinary properties, such as isospin violation \cite{delAmoSanchez:2010jr}  and radiative decays  \cite{Aubert:2008ae, Aaij:2014ala} still make its structure unsolved. Many new exotic hadron candidates have been observed by decaying into final states of a charm and anticharm quarks. These states are called $XYZ$ states. The $X$ states are neutral and have positive parity quantum number and generally seen in $J/\psi + \text{pions}$ decays. Typical example is the observation of $\chi_{c1}(3872)$. The $Y$ states are neutral and have negative parity quantum number, and seen in $e^+e^-$ annihilation with or without initial state radiation. $Y(4260)$ is an example of this family \cite{Aubert:2005rm}.  The $Z$ states are mostly charged or neutral, have typically positive parity and decay into $J/\psi + \pi$, $h_c(1P) + \pi$, $\chi_c(1P) + \pi$. The charged  $Z_c(3900)$ state is the famous example of $Z$ family \cite{Ablikim:2013mio, Liu:2013dau}.  Observation of pentaquark states in 2019 made exotic hadrons more interesting \cite{Aaij:2015tga}.

The physics of exotic hadrons is a thorough piece of research which involves both short and long distance behaviors of QCD. At one side, increasing values of radius $r$ are needed when considering the spectroscopy. This is the place where nonperturbative effects are at the stage. On the other side hard processes, such as decays, occur at short distances, i.e., short values  of radius $r$. This is the place where perturbative effects are at the stage. There are two different regimes which make theoretical predictions difficult. Many theoretical and phenomenological models are being studied to understand and interpret these exotic states such as lattice QCD,  dynamically generated resonances, QCD sum rules, coupled channel effects and nonrelativistic effective field theories (see Ref. \cite{Ghalenovi:2020zen} and references therein).

Generally two pictures are taken into account in the approaches mentioned above: molecular and compact tetraquark pictures. Hadronic molecules are loosely bounded systems together by the exchange of pions and other light mesons. This scenario has received a lot of interest due to the masses of several $XYZ$ hadrons are very close to the related meson-antimeson thresholds. In the case of $\chi_{c1}(3872)$ it has been suggested that if it has a binding energy of less than 200 keV with respect to the $D^0 \bar{D}^{\ast 0}$, according to the $R=1/\sqrt{2\mu E_B}$, where $\mu$ is the reduced mass of the two-hadron system and $E_B$ is the binding energy,  it  would be at least as large as 10 fm \cite{Guo:2017jvc}. From this perspective, hadronic molecules can be seen as extended objects. Compact tetraquarks are bound states of color nonsinglet diquark-antidiquarks, tightly bound by gluons, very much along the same lines as colored quark-antiquark pairs are bound into color-neutral mesons \cite{Ali:2019roi}. A diquark is a bound quark-quark $(qq)$ pair, whereas an antidiquark is a bound antiquark-antiquark $(\bar{q}\bar{q})$ pair. These pairs are colored, i.e., have non-zero color charges and can have colorless combinations which turns out to be an ansatz of tetraquark paradigm. Tetraquark configurations are not ruled out by QCD. Indeed this context opens a new window of compact hadrons in QCD, which are even more numeorus than the conventional quark-antiquark mesons.

A very recent study of LHCb present a $J/\psi$-pair invariant mass spectrum  by using pp collision data \cite{Aaij:2020fnh}. A narrow structure around 6.9 $\text{GeV/c}^2$ matching the lineshape of a resonance and a broad structure just above twice the $J/\psi$ mass are observed.  The deviation of the data from nonresonant $J/\psi$-pair production is above five standard deviations in the mass region between 6.2 and 7.4 $\text{GeV/c}^2$, covering predicted masses of states composed of four charm quarks \cite{Aaij:2020fnh}. This energy range lie well above the experimentally known range for charmonium which is in the range of 3 - 4.5 GeV and makes fully-charm tetraquark state $cc\bar{c}\bar{c}$ very interesting. The reason for this is that, the energy range of the $XYZ$ states are in the same mass range of the conventional charmonium states and this can yield a confusion on these structures \cite{Debastiani:2017msn}. The observation of a possible $cc\bar{c}\bar{c}$ structure triggered many theoretical studies \cite{Yang:2020wkh, Wan:2020fsk,Gong:2020bmg,Zhu:2020snb,Cao:2020gul,Guo:2020pvt,Zhang:2020xtb,Weng:2020jao,Faustov:2020qfm,Ma:2020kwb,Wang:2020dlo,Karliner:2020dta,Maciula:2020wri,Eichmann:2020oqt,Wang:2020wrp,Chao:2020dml,Richard:2020hdw,Becchi:2020uvq,Lu:2020cns,Jin:2020jfc}. 

In the side of $bb\bar{b}\bar{b}$ structures, there is no observation up to now. An experimental study claimed  the existence of a full-bottom tetraquark states $bb\bar{b}\bar{b}$, with a global significance of 3.6 $\sigma$ and a mass around 18.4 GeV, almost 500 MeV below the threshold of $\Upsilon \Upsilon$ \cite{durgut}. However, LHCb Collaboration presented an intriguing analysis looking for the exotic $bb\bar{b}\bar{b}$ tetraquark in the $\Upsilon(1S) \mu^+ \mu^- $ final state and announced that no observation was made  \cite{Aaij:2018zrb}. Prior to and after these experimental studies, the possible existence of $bb\bar{b}\bar{b}$ state was investigated on the theoretical basis \cite{Heller:1985cb, Berezhnoy:2011xn,Chen:2016jxd,Karliner:2016zzc,Bai:2016int,Eichten:2017ual,Wang:2017jtz,Esposito:2018cwh,Chen:2020xwe}.

Motivated with this charm sector observation and a possible open window for fully-bottomed tetraquark states, in the present work we will use a nonrelativistic model to  study tetraquarks as composed of diquarks and antidiquarks, which interact much like ordinary quarkonia. Owing to the masses of the valence degree of freedoms, fully-heavy four quark states can be investigated by nonrelativistic approach. We calculate mass spectra of fully-heavy tetraquark systems assuming that they are composed of doubly heavy quark ($QQ$) and antidiquark ($\bar{Q}\bar{Q}$). Choosing this configuration is not just an assumption. Firstly, in light quark systems, the binding mechanism is maintained by the light-meson (such as pion, $\pi$) and gluon exchange.   The binding mechanism in fully-heavy systems is probably dominated by the gluon-exchange forces since the typical gluon mass scale is $m_g \sim 0.5 ~ \text{GeV}$. This mass value is much lighter than the possible force carriers of heavy-mesons that could be exchanged between the heavy diquark ($QQ$) and antidiquark ($\bar{Q}\bar{Q}$) of tetraquark structure. Secondly, there is some evidence of diquark clustering in baryons. In 2017, the LHCb Collaboration reported the observation of a doubly charmed and doubly charged baryon, $\Xi_{cc}^{++}$ a $ccu$ state where the charm diquark may play a role in the structure, has lead to further attention on heavy-quark systems as the description of exotic hadrons \cite{Aaij:2017ueg}.

This paper is organized as follows: In Section \ref{sec:level2}, we describe nonrelativistic potential model of this work. In Section \ref{sec:level3}, the masses of diquark/antidiquark and tetraquark systems are calculated. Detailed comparisons of diquark and tetraquark masses with previous studies within different approaches are given. Section \ref{sec:level4} is reserved for conclusion and summary of the obtained results. 

\section{\label{sec:level2}Potential Model}
The fundamental assumption in the quark potential model is that, if one integrate out the gluon fields in the QCD action, he/she can hopefully obtain an effective Hamiltonian with suitable potentials which describes the physics of hadrons fairly good. The quark masses in the resulting effective Hamiltonian might not be the same quark masses in the QCD Lagrangian, that is why they are named as constituent quark masses. 

The use of quark potential models to describe the energy spectra of mesonic and baryonic systems gave reliable results. Phenomenological models with a simple relativistic kinetic energy term plus a scalar potential term which incorporates the so called linear confinement plus a term related to short distance which incorporates color-Coulomb interaction stemmed from QCD, give good results and descriptions of the observed spectra of both heavy and light quark mesons and baryons.

In the present work, we have considered the following potential:
\begin{equation}
V(r)=V_H(r) + V_C(r) + V_0. \label{totalpot}
\end{equation}
Here, $V_H$ denotes the Hulthen potential:
\begin{equation}
V_H(r)=-\frac{h}{\text{exp}(br)-1}, ~b >0, ~ h \geq 0,
\end{equation}
where $h$ is the coupling strength and $b$ is the range. Hulthen potential is extensively used in atomic, molecular, plasma and mathematical physics \cite{malli1980,lindhard1986,jia2000,candemir2013,bahar2016,
karayer2017}, chemical physics \cite{pyyko1975,ikhdair2007,amlan2015}, nuclear physics \cite{lam1971,rlh1985,Laha:2015ffa}, and particle physics \cite{Bhoi:2013wba,Hosseinpour:2016ujj,aktas2018}.

In the limit $b \to 0$, the Hulthen potential approaches from above the Coulomb-like potential 
\begin{equation}
V_H(r)= -\frac{h}{br} \approx V_{\text{Coulomb}}(r)=-\frac{\kappa}{r}.
\end{equation}
In other words, the Hulthen potential behaves like the Coulomb potential as $r \to 0$ but decreases exponentially in the asymptotic region when $r$ is sufficiently large. $V_C$ is the confining part of the potential with
\begin{equation}
V_C(r)=cr,
\end{equation}
where $c$ is a constant. $V_0$ is also a constant which would act as a zero-point energy. The parameters for this potential are given in the Table \ref{tab:table1} \cite{Bhaghyesh:2011zza}:

\begin{table}[H]
\caption{\label{tab:table1}Potential model parameters.}
\begin{ruledtabular}
\begin{tabular}{lc}
Parameter&Numerical value\\
\hline
$h$ & $0.20 ~ \text{GeV}$  \\
$b$ & $0.4 ~ \text{GeV}$ \\
$c$ & $0.193 ~ \text{GeV}^2$ \\
$V_0$ & $-0.223 ~ \text{GeV}$ \\
$m_c$ & $1.4 ~ \text{GeV}$ \\
$m_b$ & $4.812 ~ \text{GeV}$ \\
$\alpha_s(m_c^2)$ & 0.37 \\
$\alpha_s(m_b^2)$ & 0.26 \\
\end{tabular}
\end{ruledtabular}
\end{table}

\begin{figure}[H]
\includegraphics[width=3.4in]{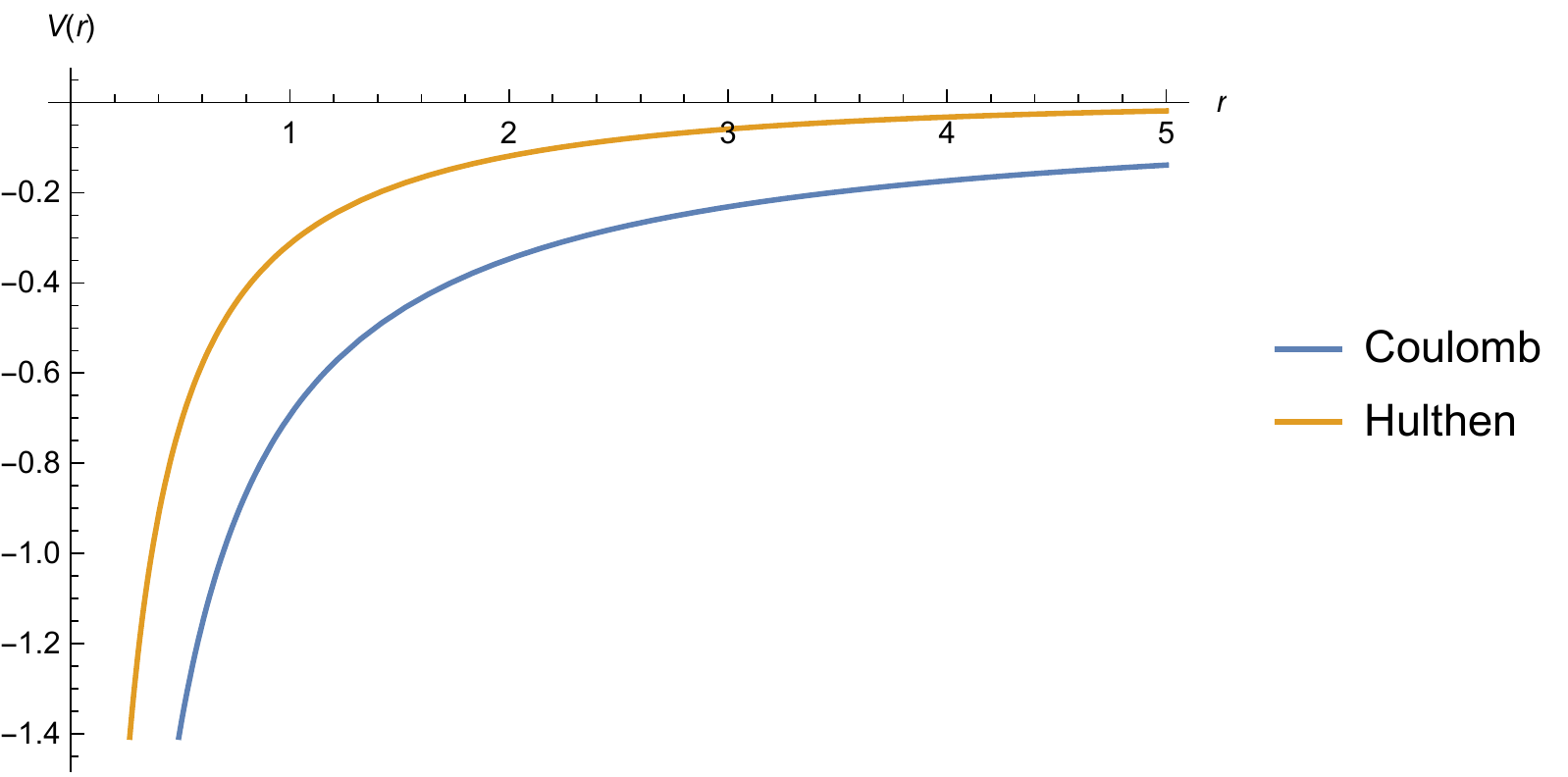}% Here is how to import EPS art
\caption{\label{fig:1} Comparison of Coulomb potential and Hulthen potential for large $r$. $V(r)$ is in GeV and $r$ in $\text{fm}$ units.}
\end{figure}

To see whether or not $V(r)$ (Eq. \ref{totalpot}) coincides with the properties of Cornell potential exhibit, we can plot them. At first step, it can be shown that  for large $r$, the Hulthen potential approaches to zero faster than the Coulomb-like potential, $V(r)=-\frac{4}{3}\frac{\alpha_s}{r}$  with $\alpha_s=0.5202$ \cite{Debastiani:2017msn}. This can be seen in Figure \ref{fig:1}.

The choice of Hulthen potential could be explained as follows. As can be seen from Figure \ref{fig:1},  the contribution of large $r$ value of Hulthen potential is less than the Coulomb potential as $r$ increases. This is due to the nature of short range potentials. Furthermore, the general theory of scattering is not immediately applicable to the Coulomb potential case because it decreases too slowly as the distance increases, which can be seen in Figure \ref{fig:1}. The Coulomb potential falls off so slowly that it continues to influence the particles even as they move apart \cite{Taylor:1983}. In order to apply the general theory of scattering to the Coulomb potential case, some modifications of the Coulomb potential have been considered. For example replacing Coulomb potential by the Yukawa potential and subsequently making the exponential factor approach unity is one example of this modification. Hulthen potential is a good approximation of Yukawa potential, $V(r)=-V_0 \frac{e^{-\alpha r}}{r}$. Furthermore, Hulthen potential combined with a linear term, $\sim r^n$, can be a theoretical playground for the different forms of linear part of full potentials.

In addition to above discussion, the spectrum generated by the Hulthen potential could be investigated. Since this is done in literature excessively, it will be good to mention the well-known operator inequality. It was shown in \cite{Lucha:2014mdw} that
\begin{equation}
V_{\text{Coulomb}}(r)=-\frac{\kappa}{r} \leq -\frac{h}{br} \leq -\frac{h}{\text{exp}(br)-1}=V_H(r),
\end{equation} 
for $\frac{h}{b} \leq \kappa$.
This operator inequality between the Coulomb and Hulthen potentials is shared by the corresponding Hamiltonians and thus carries over to the entailed (ground state) energy eigenvalues.

\begin{figure}[H]
\includegraphics[width=3.4in]{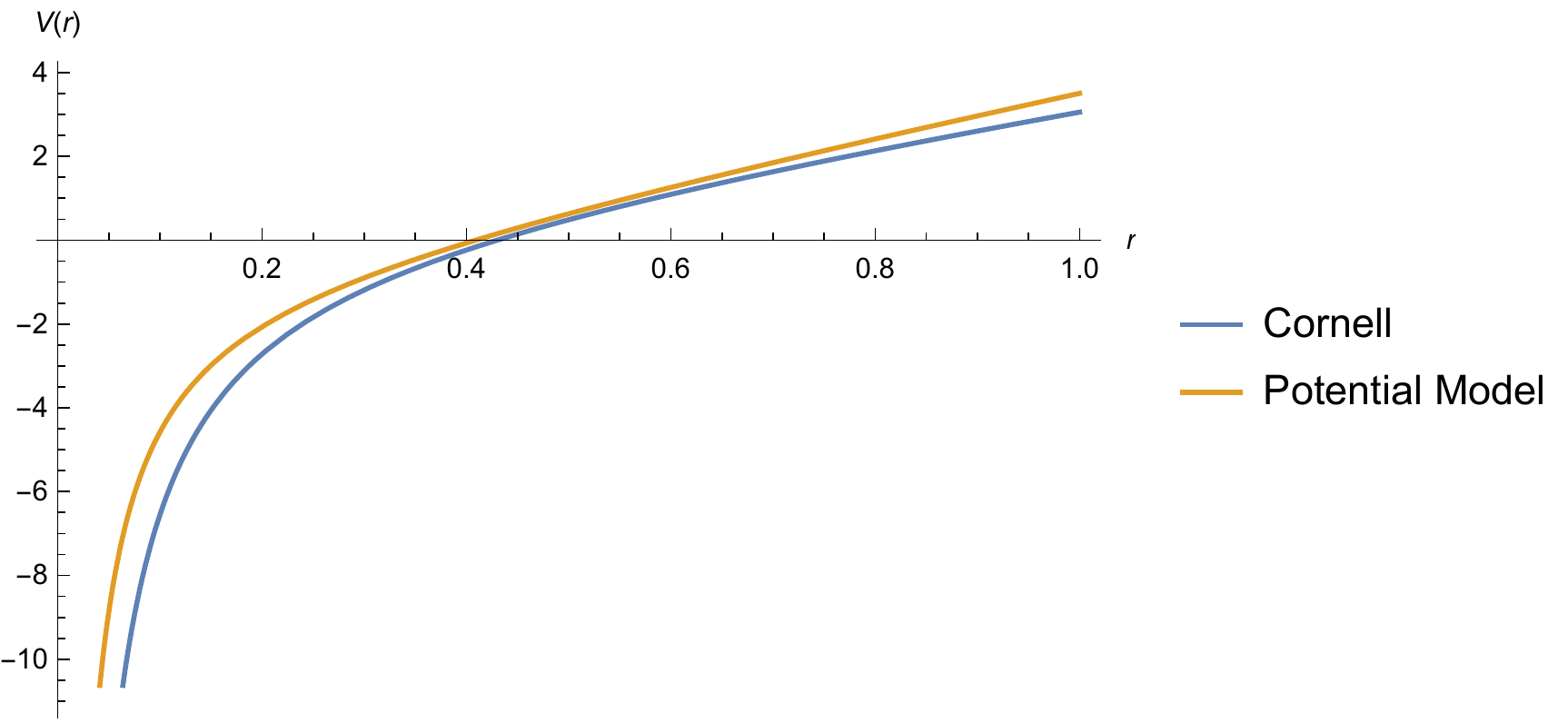}% Here is how to import EPS art
\caption{\label{fig:2} Comparison of Cornell potential and potential model of this work. $V(r)$ is in GeV and $r$ in $\text{fm}$ units.}
\end{figure}

On the other side, the plots of Cornell potential $V(r)=-\frac{4}{3}\frac{\alpha_s}{r}+br$ with $\alpha_s=0.5202$ and $b=0.1463~ \text{GeV}^2$ \cite{Debastiani:2017msn}, and potential model of this work are given in Figure \ref{fig:2}. 

It can be seen from Figure \ref{fig:2} that in the range of $0< r<1 ~ \text{fm}$, both potentials have similar behaviour. In literature, all the phenomenological potentials have almost similar behaviour in the range of confinement, $r< 1 ~ \text{fm}$. This is the characteristic interval of heavy quarkonium systems, charmonium $c\bar{c}$ and bottomonium $b\bar{b}$. At its present status, it is not possible to obtain  an exact  interquark potential in the whole range of distances from the first principle of QCD. 

A nonrelativistic approach with static potentials can be reasonable under the condition that the kinetic energy is much less than the rest masses of the constituents, which is usually the case when considering heavy quark bound states. Another advantage of assuming tetraquarks are composed from the heavy diquark ($QQ$) and heavy antidiquark ($\bar{Q}\bar{Q}$) is that it significantly simplifies calculations, since the four-body problem of a four quark structure is very complicated. In the context of diquark-antidiquark configuration, the problem turns out to be a two two-body problem. In a two-body problem which involves a central potential, it is legitimate
to work in the center-of-mass frame (CM), where one can use spherical coordinates to separate the radial and angular parts of the wave function. In terms of the reduced mass, let $\mu \equiv m_1 m_2/(m_1 +m_2)$ where $m_1$ and $m_2$ are the constituent masses of quark 1 and quark 2, respectively. Since we are dealing equal masses of quarks, $m \equiv m_1 = m_2$ yields reduced mass as $\mu=m/2$. Hence, the time-independent radial Schrödinger equation can be written as
\begin{equation}
\left[\frac{1}{2\mu} \left( -\frac{d^2}{dr^2}+\frac{L(L+1)}{r^2}  \right) + V(r)    \right]y(r)=Ey(r), \label{schr}
\end{equation}
with the orbital quantum number $L$ and the energy eigenvalue $E$. In order to account full spectra and splittings in energy between states with different quantum numbers, one can include spin-dependent terms to the potential. Due to the two nature of energies (high distances and short distances), it is not clear how the spin dependent forces change with distance. Assuming that spin dependent forces stem from short distance potential, observed hadron masses can be explained within well accuracy. For more about the nature of these spin dependent forces see Refs. \cite{DeRujula:1975qlm,Eichten:1980mw}. The spin-dependent parts of the potential are sensitive to the Lorentz structure of the interquark potential. Hence, one has to split the total potential as the sum of Lorentz vector $V_v(r)$ and Lorentz scalar $V_s(r)$.  Based on the Breit–Fermi interaction for one-gluon-exchange \cite{Lucha:1991vn,Lucha:1995zv,Voloshin:2007dx}, the following spin dependent terms can be added to the potential. For equal masses $m=m_1=m_2$, with $m$ being the constituent mass of the two-body problem (charm quark, bottom quark, or diquark) spin-spin interaction can be defined as
\begin{equation}
V_{SS}=\frac{2}{3m^2}   \nabla^2 V_v(r) \textbf{S}_1 \cdot \textbf{S}_2 \label{vss}.
\end{equation}
This interaction is related to the $S-$wave splittings. This term has no effect for $\ell \neq 0$ states. The expectation value of the operator for the spin-spin interaction can be calculated in terms of the spin quantum numbers by using 
\begin{equation}
 \langle \textbf{S}_1 \cdot \textbf{S}_2 \rangle =  \langle \frac{1}{2}( S^2-S_1^2-S_2^2) \rangle
\end{equation}
where $S$, $S_1$, and $S_2$ is the total spin, the spin of quark 1, and the spin of quark 2, respectively. 

The contribution of spin-dependent terms can be calculated by writing total potential Eq. (\ref{totalpot}) as sum of Lorentz vector structure and Lorentz scalar structure:
\begin{equation}
V(r)=V_v(r) + V_s(r).
\end{equation}
The Lorentz structures are well known, for example, of Cornell potential but it is not clear in the Eq. (\ref{totalpot}). To account these Lorentz structures, it was assumed that the terms in  Eq. (\ref{totalpot}) have partly vector and partly scalar \cite{Bhaghyesh:2012zz}. In our formalism, as an alternative we use a different manner. The spin-spin correction to the nonrelativistic potential can be obtained as \cite{Kiselev:1994rc}:
\begin{eqnarray}
\langle  V_{SS} \rangle &=&\frac{8 \pi}{3} \frac{\kappa_s \alpha_s}{m^2} \int \Psi^\ast (\textbf{r}) \Psi (\textbf{r}) \delta(\textbf{r}) \langle \textbf{S}_1 \cdot \textbf{S}_2 \rangle d^3r  \nonumber \\
&=& \frac{8 \pi}{3} \frac{\kappa_s \alpha_s}{m^2} \vert \Psi(0) \vert^2 \langle \textbf{S}_1 \cdot \textbf{S}_2 \rangle. \label{envss}
\end{eqnarray}
If the spin-spin interaction was treated as a first-order perturbation
without the Gaussian smearing, as it is the case in this work, it would be proportional to modulus of the wavefunction at the origin, $\vert \Psi(0) \vert ^2$. Since spin-spin interaction only occurs in $S-$wave, only $S-$wave states (i.e., orbital angular momentum $\ell=0$) have non-zero value of the wavefunction at the origin. Therefore for $S-$wave state we have \cite{Lucha:1991vn}:
\begin{equation}
 \vert \Psi(0) \vert^2= \vert Y_0^0 (\theta, \phi) R_{n, \ell}(0) \vert^2 =\frac{\vert  R_{n, \ell}(0) \vert^2}{4 \pi}.
\end{equation}
$\vert  R_{n, \ell}(0) \vert^2$ can be obtained directly from the numerical calculations and is related to the radial potential as
\begin{equation}
\vert \Psi(0) \vert^2= \frac{\mu}{2 \pi} \langle \frac{d}{dr} V(r)\rangle \Rightarrow \vert  R_{n, \ell}(0) \vert^2=2 \mu \langle \frac{d}{dr} V(r)\rangle,
\end{equation}
where $\mu$ is the reduced mass, $\mu=m_1m_2/(m_1+m_2)$. Then it is possible to replace  $\vert \Psi(0) \vert^2$ in Eq. (\ref{envss}) with $\vert  R_{n, \ell}(0) \vert^2$.

In QCD, the coupling constant $\alpha_s$ is not actually a ``constant". It changes according to the energy scale of each bound state. Therefore it changes with respect to energy scale and that is why it is called ``running" coupling constant. We have used the nonrelativistic model of Ref. \cite{Bhaghyesh:2011zza} where $\alpha_s$ is constant and has different values for fitting quarkonium spectra. Choosing $\alpha_s$ constant is a common approach in many of the nonrelativistic quark potential models.

By solving Schrödinger equation, one can obtain the  energy eigenvalue $E$. The result will depend  on the number of nodes of the wave function $n$ (or principal quantum number $N=n+1$) and the orbital angular momentum number $\ell$. The mass calculated in this case is called spin-averaged mass. If one include spin-dependent terms in the potential, the result will also depend on the total spin $S$ of the constituents spins $S_1$ and $S_2$. The Schrödinger equation has no analytical
solution for the potential given in this work. So we solve it numerically inspired from Ref. \cite{Lucha:1998xc}. 

After having solved Schrödinger equation, the mass $M$ of a particular state can be written as
\begin{equation}
M=2m_{q/d}+E+ \langle V_{ss} \rangle,
\end{equation} 
where $m_q$ is the mass of corresponding quark/antiquark and  $m_d$ is the mass of corresponding diquark/antidiquark. We have considered the diquarks and antidiquarks as constituents of the tetraquarks to predict the tetraquark masses. Therefore, we first calculate the diquark masses and then tetraquark masses.

\section{\label{sec:level3}Numerical Analysis and Discussion}
To study tetraquark systems in terms of diquark states, we will elaborate four-particle system as two two-body systems, i.e., diquark-antidiquark system. We assumed that the interaction between the diquarks and antidiquarks is to be effectively the same for ordinary quarkonia. In this picture, a diquark and antidiquark interact as a whole which means interactions between quarks from a diquark system ($QQ$) with antiquarks from an antidiquark system ($\bar{Q}\bar{Q}$) are not considered. Since quark and antiquark masses are same one does not have to calculate antidiquark masses separately. We first calculate diquark masses (antidiquark masses at the same time) and use these diquark (antidiquark) masses to calculate tetraquark masses in diquark-antidiquark picture. The motivation for this factorization is the color structure of diquarks.

Hadrons can exist only when their total color charge of constituent quarks are zero. Technically, this means that every naturally occuring hadron is a color singlet under the group symmetry SU(3). A diquark is  composed of two quarks ($qq$) whereas conventional quark-antiquark (also called quarkonium) is composed of ($q\bar{q}$). The difference between quark-quark systems and quark-antiquark systems is mainly due to the color structure.

Regarding the Cornell potential, $V(r)=\kappa \frac{\alpha_s}{r} + br$, where $\kappa$ is the color factor, $\alpha_s$ is the QCD fine structure constant and $b$ is string tension and related to the strength of the confinement, the color factor change in the diquark configuration. The SU(3) color symmetry of QCD implies that, when we combine a quark and an antiquark in the fundamental color representation, we obtain $\vert q \bar{q} \rangle:  \textbf{3}  \bigotimes  \bar{\textbf{3}}= \textbf{1} \bigoplus \textbf{8} $. This representation gives the color factor for the color singlet as $\kappa=-4/3$ of the quark-antiquark system. When we combine two quarks in the fundamental color representation, it reduces to $\vert q q \rangle:  \textbf{3}  \bigotimes  \textbf{3}= \bar{\textbf{3}} \bigoplus \textbf{6} $, a color antitriplet  $\bar{\textbf{3}}$ and a color sextet $\textbf{6}$. In a similar way, combining two antiquarks reduces to $\vert \bar{q} \bar{q} \rangle:  \bar{\textbf{3}}  \bigotimes  \bar{\textbf{3}}= \textbf{3} \bigoplus \bar{\textbf{6}} $, a triplet $\textbf{3}$ and antisextet $\bar{\textbf{6}}$. Accordingly,  combining an antitriplet diquark and a triplet antidiquark reduce to $\vert \left[ qq \right]-\left[ \bar{q} \bar{q} \right] \rangle : \textbf{3} \bigotimes \bar{\textbf{3}}= \textbf{1} \bigoplus \textbf{8}$, and form a color singlet for which the one-gluon exchange potential is attractive. The antitriplet state has a color factor $\kappa=-2/3$ which is attractive whereas the sextet state  has a color factor $\kappa=+1/3$ which is repulsive. Therefore we will only consider diquarks in the antitriplet color state. This conclusion was achieved, for example by Ref. \cite{Wu:2016vtq}, in which it is shown that for single-flavor tetraquarks, only the antitriplet diquarks can build pure states. 

This difference in color structure of the quark-antiquark and quark-quark systems make possible to extend the quark-antiquark model to be valid in the model of quark-quark system, just by changing the color factor $\kappa$ and the string tension $b$. Changing color factor $\kappa=-4/3$ (for quark-antiquark system in color singlet state) to $\kappa=-2/3$ (quark-quark system in the antitriplet color state) is equivalent of introducing a factor of 1/2
in the Coulomb part of  the Cornell potential for the conventional quark-antiquark system. This factor should be taken as a global factor since it comes from the color structure of the wave function. Therefore the string tension should also be divided by a factor of 2. The general rule for diquark potential from quark-antiquark potential is making $V_{qq}=V_{q\bar{q}}/2$. This conclusion was done in different tetraquark models \cite{Debastiani:2017msn,Ebert:2007rn,Lu:2016zhe,Lundhammar:2020xvw}. So we also divide our potential (Eq. \ref{totalpot}) by a factor of 2 for obtaining diquark spectra.

\subsection{\label{sec:level31}  Diquarks}
We now present our numerical calculations for masses of diquarks composed of charm ($c$) and bottom ($b$) quarks, which are also equivalent for antidiquark cases in this work. We assume that information in the spin-dependent interaction described in Eq. (\ref{vss}) is inherited when changing from quark-antiquark system to quark-quark system. In other words, spin-spin interaction is encoded in the
diquarks.  Similar assumptions were made in \cite{Debastiani:2017msn,Lundhammar:2020xvw,Maiani:2014aja} and gave reliable results. In the present work, we choose the attractive color antitriplet state which is antisymmetric in the color wavefunction. In order to maintain Pauli exclusion principle, the diquark total spin must be 1, $S=1$. So the total wavefunction of the diquark will be antisymmetric.

Based on the previous arguments, the results for the ground and first radially excited diquark masses are
presented in Table \ref{tab:table3}.

\begin{table}[H]
\caption{\label{tab:table3}Results for the four diquark masses for $cc$ and $bb$. Corresponding diquark masses are same as $cc$ and $bb$, respectively. All results are in MeV unit. }
\begin{ruledtabular}
\begin{tabular}{lcr}
Diquark&$N ^{2S+1}L_J$&Mass\\
\hline
$cc$ \\ \hline 
& $1^3S_1$ & 3114  \\
& $2^3S_1$ & 3443 \\
\hline 
$bb$ \\ \hline 
& $1^3S_1$ & 9792  \\
& $2^3S_1$ & 10011 \\
\end{tabular}
\end{ruledtabular}
\end{table}
Comparison of ground and first radially excited states for the $cc$ diquark masses of earlier works are shown in Table \ref{tab:table4}.

\begin{table}[H]
\caption{\label{tab:table4}Comparison of ground and first radially excited states for $cc$ diquark masses. All results are in GeV unit. }
\begin{ruledtabular}
\begin{tabular}{cccc}
State&This work&\cite{Debastiani:2017msn} & \cite{Kiselev:2002iy}\\
\hline
 $1S$ & 3.114 & 3.133& 3.13 \\
 $2S$ & 3.443 & 3.456 &3.47 \\
\end{tabular}
\end{ruledtabular}
\end{table}
As can be seen from Table \ref{tab:table4}, masses for ground and first radially excited $cc$ diquark states are in good agreement. We note that Ref. \cite{Debastiani:2017msn} used a Cornell potential with nonrelativistic  framework and Ref.   \cite{Kiselev:2002iy} used a nonrelativistic QCD motivated potential. We also compare ground states masses of $cc$ diquark with the available theoretical studies in Table \ref{tab:table5}.

\begin{table}[H]
\caption{\label{tab:table5}Comparison of $cc$  diquark masses of this work and results from other works. All diquarks are considered to be in the ground state $N ^{2S+1}L_J=1^3S_1$. All results are in MeV.}
\begin{ruledtabular}
\begin{tabular}{cc}
Reference&Mass\\
\hline 
This work & 3114 \\
\cite{Debastiani:2017msn} & 3133 \\
\cite{Faustov:2020qfm} & 3226 \\
\cite{Berezhnoy:2011xn} & 3130 \\
\cite{Karliner:2016zzc} & 3204 \\
\cite{Lundhammar:2020xvw}& 3128 \\
\cite{Kiselev:2002iy} & 3130 \\
\cite{Ferretti:2019zyh} & 3329 \\
\cite{Bedolla:2019noq} & 3144 \\
\cite{Esau:2019hqw} & 3510 $\pm$ 350\\
\cite{Yu:2018com} [Mod. Ia] & 3400 \\
\cite{Yu:2018com} [Mod. Ib] & 3370 \\
\cite{Yu:2018com} [Mod. IIa] & 3420 \\
\cite{Yu:2018com} [Mod. IIb] & 3370 \\
\end{tabular}
\end{ruledtabular}
\end{table}
As can be seen from Table \ref{tab:table5}, $cc$ diquark masses lie in the range of $3.1-3.5 ~ \text{GeV} $. Differences are due to the models of the references. In the present work, diquark masses depend on the parameters of the potential model, which were given in Section \ref{sec:level2}. Compared with the values of the reference works, the results deviate with at most 400 MeV. Our result is in good agreement with the results in Refs. \cite{Debastiani:2017msn, Berezhnoy:2011xn, Lundhammar:2020xvw,Kiselev:2002iy} which used nonrelativistic quark model. Ref. \cite{Faustov:2020qfm} used  diquark-antidiquark picture in the framework of the relativistic quark model based on the quasipotential approach. In their model, quark-quark and diquark-antidiquark interactions in the potential are constructed similar to the mesons and baryons. Their $cc$ diquark mass is approximately 100 MeV heavier than our result. In Ref. \cite{Karliner:2016zzc}, diquark masses are calculated by a framework based on existing empirical information about mesons and baryons. The predicted value is approximately 100 MeV higher than our result. In Ref. \cite{Ferretti:2019zyh},  diquark masses are calculated by taking into account spin-spin, spin-orbit, and tensor interactions in quark model. The predicted value of Ref. \cite{Ferretti:2019zyh} is higher by about 200 MeV compared to our result. In Ref. \cite{Bedolla:2019noq}, Schwinger-Dyson and Bethe-Salpeter equations are solved in order to obtain diquark masses by $\text{vector} \times \text{vector} $ interaction model (denoted as CI  in reference paper) in which the obtained mass agree well compared to our result.  A QCD Sum Rule study was done in Ref. \cite{Esau:2019hqw} yielding  a mass value which is consistently bigger by about 400 MeV than our result. Ref. \cite{Yu:2018com} made an exploratory study with doubly heavy baryon which is composed of a heavy diquark and a light quark and calculated diquark masses in the Bethe-Salpeter (BS) formalism. They used two different parameters sets with confining parameter $\kappa^\prime$ and coupling strength $\alpha_s$. They also took care of heavy quark limit. Mod. Ia refers $m_Q \to \infty$, i.e., taking the heavy quark limit while  Mod. Ib refers $m_Q \to \text{finite}$, i.e., without heavy quark limit with appropriate values of $\kappa^\prime$ and $\alpha_s$. The heavy quark limits are the same but parameters are different for Mod. IIa and Mod. IIb.  Our results are at most 300 MeV lower than the results of  Ref. \cite{Yu:2018com}.  

In Table \ref{tab:table6}, we compare $bb$ diquark masses of ground and first radially excited states of different works.
\begin{table}[H]
\caption{\label{tab:table6}Comparison of ground and first radially excited states for $bb$ diquark. All results are in GeV unit. }
\begin{ruledtabular}
\begin{tabular}{cccc}
State&This work& \cite{Kiselev:2002iy} &\cite{Gershtein:2000nx} \\
\hline
 $1S$ & 9.792 &  9.72 & 9.74 \\
 $2S$ & 10.011 & 10.01 & 10.02 \\
\end{tabular}
\end{ruledtabular}
\end{table}
It can be seen from Table \ref{tab:table6} that ground and first excited $bb$ diquark masses are in good agreement. We note that Ref. \cite{Gershtein:2000nx} used  the QCD potential of Buchmüller and Tye under nonrelativistic quark model framework.

As happened before in the $cc$ diquark case, we compare ground state masses of $bb$ diquark  with the available theoretical studies in Table \ref{tab:table7}.

\begin{table}[H]
\caption{\label{tab:table7}Comparison of $bb$  diquark masses of this work and results from other works. All diquarks are considered to be in the ground state $N ^{2S+1}L_J=1^3S_1$. All results are in MeV.}
\begin{ruledtabular}
\begin{tabular}{cc}
Reference&Mass\\
\hline 
This work & 9792 \\
\cite{Faustov:2020qfm} & 9778 \\
\cite{Berezhnoy:2011xn} & 9720 \\
\cite{Karliner:2016zzc} & 9719 \\
\cite{Lundhammar:2020xvw}& 9643 \\
\cite{Kiselev:2002iy} & 9720 \\
\cite{Ferretti:2019zyh} & 9845 \\
\cite{Bedolla:2019noq} & 9491 \\
\cite{Esau:2019hqw} & 8670 $\pm$ 690\\
\cite{Yu:2018com} [Mod. Ia] & 10070 \\
\cite{Yu:2018com} [Mod. Ib] & 10050 \\
\cite{Yu:2018com} [Mod. IIa] & 10080 \\
\cite{Yu:2018com} [Mod. IIb] & 10050 \\
\cite{Gershtein:2000nx} & 9740 \\
\cite{Anwar:2017toa} & 9850 \\

\end{tabular}
\end{ruledtabular}
\end{table}

The $bb$ diquark masses lie in the range of $8.6-10.0 ~ \text{GeV} $. This gap is bigger with respect to $cc$ diquark mass which was around 400 MeV. The reason for this could be the constituent mass of $b$ quark, which is roughly three or four times $c$ quark mass in constituent quark models.  Nonrelativistic quark model results agree well with our result \cite{Debastiani:2017msn, Berezhnoy:2011xn, Lundhammar:2020xvw,Kiselev:2002iy,Gershtein:2000nx}. Ref. \cite{Anwar:2017toa} calculated $bb\bar{b}\bar{b}$ tetraquark mass in a nonrelativistic effective field theory and in a relativized diquark model characterized by one-gluon-exchange (OGE) plus a confining potential. In order to do this, they estimated $bb$ diquark mass by binding the OGE plus a confining potential. Their result agree well with our result.

The differences of our results in $cc$ and $bb$ diquark masses apart from nonrelativistic formalisms could be as a result of the relativistic effects, spin-orbit and tensor interactions, model parameters or method uncertainties (there exist a typical 10\% uncertainty in the results of QCD Sum Rules).

\subsection{\label{sec:level32}Tetraquarks}
In tetraquark systems, we treated fully-heavy tetraquark as a two body ($QQ-\bar{Q} \bar{Q}$) system with equal masses, $m_{QQ}=m_{\bar{Q}\bar{Q}}$. We also treated them as axial-vector states composed of diquarks and antidiquarks and we assumed that the interaction between the diquarks and antidiquarks mimickes the ordinary quarkonia interaction. It should be also noted that, tetraquarks made up of quarks of the same flavor can only be made up of axial-vector diquark-antidiquark pairs because of Pauli principle.

One of the important advantages of quark model is that, it allows to study systems with various  combinations of principal quantum number, orbital angular momentum, spin and total angular momentum. Coupling of spin 1 diquark and spin 1 antidiquark produces tetraquark of total spin 0, 1 and 2 which we denote as $S_T=0,1,2$. We transferred quantum mechanic couplings into this nonrelativistic approach where total spin $S_T$ and orbital angular momentum $L_T$ couple into the total angular momentum $J_T$. Obtaining charge and parity  quantum numbers of tetraquark states is quite different than the conventional method. We obtain charge and parity quantum numbers of tetraquarks as explained in \cite{Debastiani:2017msn,Maiani:2014aja,Maiani:2004vq}. We will use the following notation
\begin{equation}
\vert T_{QQ\bar{Q}\bar{Q}} \rangle= \vert S_d, S_{\bar{d}}, S_T, L_T , J_T \rangle,
\end{equation}
where $S_d$ is the total spin of the diquark, $S_{\bar{d}}$ is the total spin of the antidiquark, $S_T$ is the total spin of the tetraquark, $L_T$ is the orbital angular momentum relative to the diquark-antidiquark system (in the two-body scheme), $J_T$ is the total angular momentum of the tetraquark comes from the coupling of $S_T$ and $L_T$. Then, charge and parity quantum numbers can be written as
\begin{eqnarray}
C_T&=&(-1)^{L_T + S_T}, \\
P_T&=& (-1)^{L_T}.
\end{eqnarray}
Since we use diquarks/antidiquarks with spin 1 in the antitriplet color configuration, the resulting tetraquark states in $S-$wave can have the following possibilities:
\begin{eqnarray}
\vert 0^{++} \rangle &=& \vert S_d=1, S_{\bar{d}}=1, S_T=0, L_T=0, J_T=0 \rangle, \\
\vert 1^{+-} \rangle &=& \vert S_d=1, S_{\bar{d}}=1, S_T=1, L_T=0, J_T=1 \rangle, \\
\vert 2^{++} \rangle &=& \vert S_d=1, S_{\bar{d}}=1, S_T=2, L_T=0, J_T=2 \rangle.
\end{eqnarray}

Considering diquarks and antidiquarks as consituents of tetraquarks, masses of the tetraquarks can be calculated using the diquark masses with $N ^{2S+1}L_J=1 ^{3}S$ in Table \ref{tab:table3} for $cc$ and $bb$ diquarks. We also show the values of $\Delta=M_{\text{tetra}}-M_{\text{threshold}}$, where $M_{\text{tetra}}$ is the tetraquark mass and $M_{\text{threshold}}$ is the mass of its lowest meson-meson threshold. A negative $\Delta$ means that the tetraquark lies below the
threshold of the fall-apart decay into two mesons and thus should be a narrow state.
Besides that a state with small positive $\Delta$ value could also be observed as a resonance since its decay rate will be suppressed by the phase space. All other states with large positive $\Delta$ values are expected to be broad and difficult to observe in the experiments. The results are presented in Table \ref{tab:table8}.

\begin{table*}
\caption{\label{tab:table8}Masses of ground, first, and second radially excited fully-heavy $cc\bar{c}\bar{c}$ and $bb\bar{b}\bar{b}$ tetraquark states. $E_{th}$ is the threshold mass of two heavy $Q\bar{Q}$ mesons. All results are given in MeV.}
\begin{ruledtabular}
\begin{tabular}{ccccccc}
  Configuration&$N ^{2S
_T+1}L_{T_{J_T}}$ &$J^{PC}$&Mass&Threshold&$E_{th}$
&$\Delta$\\ \hline 
$cc\bar{c}\bar{c}$&$1^1S_0$&$0^{++}$&6322 &$\eta_c(1S)\eta_c(1S)$& 5968 & 354\\
                  &$1^1S_0$&$0^{++}$&6322 &$J/\psi(1S)J/\psi(1S)$& 6194 & 128\\
                  &$1^3S_1$&$1^{+-}$&6354 &$\eta_c(1S)J/\psi(1S)$& 6081 & 273\\
                  &$1^5S_2$&$2^{++}$&6385 &$J/\psi(1S)J/\psi(1S)$& 6194 & 191\\ \hline
                  
                  &$2^1S_0$&$0^{++}$& 6575 &$\eta_c(1S)\eta_c(1S)$& 5968  & 607\\
                  &$2^1S_0$&$0^{++}$& 6575 &$J/\psi(1S)J/\psi(1S)$& 6194  & 381 \\
                  &$2^3S_1$&$1^{+-}$& 6609 &$\eta_c(1S)J/\psi(1S)$& 6081  & 528\\
                  &$2^5S_2$&$2^{++}$& 6639 &$J/\psi(1S)J/\psi(1S)$& 6194  &445\\ \hline 
                  
                  &$2^1S_0$&$0^{++}$& 6575 &$\eta_c(2S)\eta_c(2S)$& 7276  & -701\\
                  &$2^1S_0$&$0^{++}$& 6575&$\psi(2S)\psi(2S)$& 7372 &-797 \\
                  &$2^3S_1$&$1^{+-}$& 6609&$\eta_c(2S)\psi(2S)$& 7324 & -715\\
                  &$2^5S_2$&$2^{++}$& 6639&$\psi(2S)\psi(2S)$& 7372&-733\\ \hline

                  &$3^1S_0$&$0^{++}$& 6782 &$\cdots$&$\cdots$   &$\cdots$\\
                  &$3^3S_1$&$1^{+-}$& 6814&$\cdots$& $\cdots$ &$\cdots$\\
                  &$3^5S_2$&$2^{++}$& 6842&$\cdots$& $\cdots$&$\cdots$\\ \hline

$bb\bar{b}\bar{b}$&$1^1S_0$&$0^{++}$&19666 &$\eta_b(1S)\eta_b(1S)$& 18798 & 868\\
				  &$1^1S_0$&$0^{++}$&19666 &$\Upsilon(1S)\Upsilon(1S)$& 18920 & 746\\
				  &$1^3S_1$&$1^{+-}$&19673 &$\eta_b(1S)\Upsilon(1S)$& 18859 & 814\\
				  &$1^5S_2$&$2^{++}$&19680 &$\Upsilon(1S)\Upsilon(1S)$& 18920 &760\\ \hline
				  
				  &$2^1S_0$&$0^{++}$&19841 &$\eta_b(1S)\eta_b(1S)$& 18798 & 1043\\
				  &$2^1S_0$&$0^{++}$&19841 &$\Upsilon(1S)\Upsilon(1S)$& 18920 & 921\\
				  &$2^3S_1$&$1^{+-}$&19849 &$\eta_b(1S)\Upsilon(1S)$& 18859 & 990\\
				  &$2^5S_2$&$2^{++}$&19855 &$\Upsilon(1S)\Upsilon(1S)$& 18920 &935\\ \hline
				  
				  &$2^1S_0$&$0^{++}$&19841 &$\eta_b(2S)\eta_b(2S)$& 19998 & -157\\
				  &$2^1S_0$&$0^{++}$&19841 &$\Upsilon(2S)\Upsilon(2S)$& 20046 & -205\\
				  &$2^3S_1$&$1^{+-}$&19849 &$\eta_b(2S)\Upsilon(2S)$& 20022 & -173\\
				  &$2^5S_2$&$2^{++}$&19855 &$\Upsilon(2S)\Upsilon(2S)$& 20046 &-191\\ \hline

				  &$3^1S_0$&$0^{++}$& 20001 &$\cdots$&$\cdots$   &$\cdots$\\
                  &$3^3S_1$&$1^{+-}$& 20012&$\cdots$& $\cdots$ &$\cdots$\\
                  &$3^5S_2$&$2^{++}$& 20021&$\cdots$& $\cdots$&$\cdots$\\
\end{tabular}
\end{ruledtabular}
\end{table*}

We see from this table that the predicted masses of ground and first radially excited  $cc\bar{c}\bar{c}$ and $bb\bar{b}\bar{b}$ tetraquarks lie significantly higher than the corresponding lowest (1S-1S) meson-meson thresholds. However masses of first radially excited $cc\bar{c}\bar{c}$ and $bb\bar{b}\bar{b}$ tetraquarks  are below than their corresponding (2S-2S) meson-meson thresholds.  1S $cc\bar{c}\bar{c}$ tetraquarks are located at $\sim$ 6.3 GeV and the mass splitting  is about 30 MeV.  $J^{PC}=0^{++}$ 1S $cc\bar{c}\bar{c}$ state is about $\sim$ 300 MeV  above than $\eta_c \eta_c$ mass threshold while $\sim$ 100 MeV above than $J/\psi J/\psi$ mass threshold. This suggests that $J^{PC}=0^{++}$ $cc\bar{c}\bar{c}$ state might easily decay into $\eta_c \eta_c$ and $J/\psi J/\psi$ final states through quark rearrangements, respectively. $J^{PC}=1^{+-}$ 1S $cc\bar{c}\bar{c}$ state lies about 270 MeV above than the $\eta_c J/\psi$ mass threshold. It might easily decay into $\eta_c J/\psi$ final state through quark rearrangements. $J^{PC}=2^{++}$ 1S $cc\bar{c}\bar{c}$ tetraquark is about 200 MeV above than the mass threshold of $J/\psi J/\psi$, which suggests that it might easily decay into $J/\psi J/\psi$ final state through quark rearrangements. The decays of 1S $cc\bar{c}\bar{c}$ tetraquarks are not suppressed  dynamically or kinematically. Therefore these states should be broad and hard to be observed in experimental studies.

The previous discussion can be easily adapted to 1S $bb\bar{b}\bar{b}$ tetraquarks. The $J^{PC}=0^{++}, 1^{+-}$, and $J^{PC}=2^{++}$ 1S $bb\bar{b}\bar{b}$ tetraquarks are located at $\sim$ 19.6 GeV and the mass splitting is about 10 MeV. This result is expected since spin-spin interaction term depend on mass as $\sim 1/m^2$. As the mass increases the contribution of spin-spin interaction term becomes small. $J^{PC}=0^{++}$ 1S $bb\bar{b}\bar{b}$ tetraquark is about $\sim$ 800 MeV above than $\eta_b \eta_b$ mass threshold and $\sim$ 700 MeV above than $\Upsilon \Upsilon$ mass threshold, respectively. This significant amount suggest that it might easily decay into $\eta_b \eta_b$ and $\Upsilon \Upsilon$ final states through quark rearrangements, respectively. $J^{PC}=1^{+-}$ 1S $bb\bar{b}\bar{b}$ state lies about 800 MeV above than the $\eta_b \Upsilon$ mass threshold. This significant amount suggests that it might easily decay into $\eta_b \Upsilon$ final state through quark rearrangements. $J^{PC}=2^{++}$ 1S $bb\bar{b}\bar{b}$ tetraquark is about 700 MeV above than the mass threshold of $\Upsilon \Upsilon$, and might easily decay into  $\Upsilon \Upsilon$ final state through quark rearrangements. The decays of 1S $bb\bar{b}\bar{b}$ tetraquarks are not suppressed  dynamically or kinematically. Therefore these states should be broad and difficult to be observed in experimental studies. This conclusion about 1S $bb\bar{b}\bar{b}$ tetraquarks agree with current experimental data of LHCb \cite{Aaij:2018zrb} and CMS \cite{Sirunyan:2020txn} Collaborations. No narrow beautiful tetraquarks in the $\Upsilon(1S)$-pair production were observed in these experiments. Indeed, a lattice QCD study was performed to investigate the existence of a $bb\bar{b}\bar{b}$ tetraquark bound state by using the first-principles lattice nonrelativistic QCD methodology \cite{Hughes:2017xie}. They studied $S$-wave  $bb\bar{b}\bar{b}$ tetraquark in three channels $0^{++}$, $1^{+-}$ and $2^{++}$,  and found no evidence of a QCD bound tetraquark below the lowest noninteracting thresholds in the channels. A similar conclusion was made in \cite{Esposito:2018cwh} by using a new compact tetraquark model. They found that the decay modes of fully-$b$ with $J^{PC}=0^{++}$ tetraquarks may be difficult to access in experiments. In Ref. \cite{Becchi:2020mjz}, a quantitative analysis of the  $bb\bar{b}\bar{b}$ tetraquark decays into 4 muons, $B^+B^-$, $B^0 \bar{B}^0$, and $B^0_s \bar{B}^0_s$ channels was made. They concluded that the 4 muon (4$\mu $) signal for the $J^{PC}=0^{++}$ ground state of $bb\bar{b}\bar{b}$ is likely to be too small for the upgraded LHCb, but there is a hope for the $J^{PC}=2^{++}$ $bb\bar{b}\bar{b}$ tetraquark. A similar study with considering different channels  was done in \cite{Goncalves:2021ytq}. They studied  the production of the fully-heavy tetraquarks by focusing on the $\gamma \gamma \rightarrow {\cal{Q}}{\cal{Q}}$  (${\cal{Q}} = J/\psi,\, \Upsilon$) subprocess  mediated by the $T_{4Q}$ resonance in the $s$-channel. They found that experimental study of this process is feasible and can be used  to explore the existence and properties of the $T_{4c}(6900)$ and $T_{4b}(19000)$ states.

The obtained mass values of 2S $cc\bar{c}\bar{c}$ and $bb\bar{b}\bar{b}$ tetraquarks are significantly above than the fall-apart decays to the lowest allowed (1S-1S) two-mesons. These states should be broad and difficult to be observed experimentally.

On the other hand,  the predicted masses of radially excited 2S $cc\bar{c}\bar{c}$ and $bb\bar{b}\bar{b}$ tetraquarks are lower than their corresponding thresholds. The $J^{PC}=0^{++}, 1^{+-}$, and $J^{PC}=2^{++}$ 2S $cc\bar{c}\bar{c}$ tetraquarks are located at $\sim$ 6.6 GeV and  mass splitting is about 30 MeV. The $J^{PC}=0^{++},1^{+-}$, and $J^{PC}=2^{++}$ 2S $bb\bar{b}\bar{b}$ tetraquarks are located at $\sim$ 19.8 GeV and  mass splitting is about 10 MeV. Very recently, the LCHb Collaboration  observed a broad structure at around $6.2$-$6.8$ GeV  which is very close to twice the $J/\psi$ mass threshold,  a narrow structure around 6.9 GeV, and a vague structure around 7.2 GeV \cite{Aaij:2020fnh}. We have several candidates of excited 2S $cc\bar{c}\bar{c}$ tetraquarks in the $6.2$-$6.8$ mass range for the broad structure observed by LHCb Collaboration \cite{Aaij:2020fnh}. Narrow structure around 6.9 GeV might be orbital excitations of 2S or 3S $cc\bar{c}\bar{c}$ tetraquarks. A recent study about LHCb of $J/\psi$ pairs was performed by using coupled channel methods in two channels $J/\psi J/\psi$ and $\psi(2S)J/\psi$ and in three channels $J/\psi J/\psi$, $\psi(2S)J/\psi$, and $\psi(3770)J/\psi$ \cite{Dong:2020nwy}. They found that in both channels, there is a hint of an existence of a near-threshold in the $J/\psi J/\psi$ system with $J^{PC}=0^{++}$ or $J^{PC}=2^{++}$ quantum numbers having a mass around 6.2 GeV which they refer this state as $X(6200)$. Our results of $J^{PC}=0^{++}$ and $J^{PC}=2^{++}$ $cc\bar{c}\bar{c}$ tetraquarks are $\sim$ 100 MeV above than their prediction. In Ref. \cite{Liu:2020tqy}, it was suggested that the structure around 7.2 GeV which was named as $X(7200)$ can be related to $\chi_{c1}(3872)$ (formerly known as $X(3872)$) via heavy anti-quark di-quark symmetry. Orbital angular momentum addition to our calculation for 3S $cc\bar{c}\bar{c}$ tetraquarks could yield a candidate for $X(7200)$.

The obtained mass values of 2S $cc\bar{c}\bar{c}$ and $bb\bar{b}\bar{b}$ tetraquarks are significantly above than the thresholds of decays to the lowest (1S-1S) two-mesons. These states should be broad and are difficult to observe experimentally.

We compare our predictions for the masses of fully-heavy tetraquarks with the results of previous calculations. At first step we compare results of some works which considered excited fully-heavy tetraquarks. The results can be seen in Table \ref{tab:table9}. 
\begin{table*}
\caption{\label{tab:table9}Comparison of calculated fully-heavy tetraquark $cc\bar{c}\bar{c}$  and $cc\bar{c}\bar{c}$  $bb\bar{b}\bar{b}$  masses for ground (1S) and radial excited states  (2S, 3S) with quantum numbers $J^{PC}=0^{++},1^{+-}$ and $2^{++}$. All results are in MeV.}
\begin{ruledtabular}
\begin{tabular}{cccccccccc}
 $J^{PC}$&\multicolumn{3}{c}{$0^{++}$}&\multicolumn{3}{c}{$1^{+-}$}&\multicolumn{3}{c}{$2^{++}$} \\   \cline{2-4} \cline{5-7} \cline{8-10}
State & $1S$ & $2S$& $3S$ &$1S$ &$2S$ &$3S$ & $1S$ &$2S$ & $3S$ \\ \hline
$cc\bar{c}\bar{c}$ &  & &  & & & &  & &  \\
 This work&6322&6575 &6782&6354 &6609&6814& 6385 & 6639 &6842 \\
 Ref. \cite{Zhao:2020nwy}&6476& 6908 &7296 &6441 &6896&7300& 6475 & 6921 &7320 \\
 Ref. \cite{Bedolla:2019zwg}&5883& 6573 &6948 &6120 &6669&7016& 6246 & 6739 &7071 \\ \hline
 
 $bb\bar{b}\bar{b}$ &  & &  & & & &  & &  \\
 This work&19666&19841 &20001&19673 &19849&20012& 19680 & 19855 &20021 \\
 Ref. \cite{Zhao:2020nwy}&19226& 19583 &19887 &19214 &19582&19989& 19232 & 19594 &19898 \\
 Ref. \cite{Bedolla:2019zwg}&18748& 19335 &19644 &18828 &19366&19665& 18900 & 19398 &19688 \\ 
 
\end{tabular}
\end{ruledtabular}
\end{table*}

In Table \ref{tab:table9},  Ref. \cite{Zhao:2020nwy} calculated the masses and wave functions of the exotic hadron states $cc\bar{c}\bar{c}$ and $bb\bar{b}\bar{b}$  solving the four-body Schrödinger equation by including Cornell potential and spin-spin interaction in vacuum and strongly interacting matter in quark model. Ref. \cite{Bedolla:2019zwg} used a relativized diquark model Hamiltonian built with relativistic kinetic energies, a one gluon exchange potential and linear confinement, in order to obtain masses of $cc\bar{c}\bar{c}$ and $bb\bar{b}\bar{b}$, and some other systems. Our results are compatible compared to reference studies. The reason of deviances in the results should be the different models used in the studies. At the second step, we compare our  results for the ground states of fully-heavy tetraquarks of available studies. The results can be seen in Table \ref{tab:table10}.

\begin{table*}
 \caption{Comparison of theoretical predictions for the ground state masses
 of the $cc\bar{c}\bar{c}$ and $bb\bar{b}\bar{b}$ tetraquarks. All results are in MeV.}
\label{tab:table10}
\begin{ruledtabular}
\begin{tabular}{ccccccc}
Reference & \multicolumn{3}{c}{$cc \bar c \bar c$}& \multicolumn{3}{c}{$bb \bar b \bar b$}\\
\cline{2-4} \cline{5-7} & $0^{++}$ & $1^{+-}$ & $2^{++}$ & $0^{++}$ & $1^{+-}$ & $2^{++}$\\
\hline
 \centering{This work} & 6322 & 6354 & 6385 & 19666 & 19673 & 19680 \\
%\cline{2-5}
  \centering{\cite{Debastiani:2017msn}} & 5969& 6021 & 6115  & $\cdots$ & $\cdots$ & $\cdots$ \\
  \centering{\cite{Weng:2020jao}}&6271& 6231 &6287 &18981 &18969 & 19000\\ 
%\cline{2-5}
 \centering{\cite{Faustov:2020qfm}} & 6190  & 6271 & 6367 & 19314 & 19320 & 19330 \\
   \centering{\cite{Lu:2020cns}}&6542 &6515 &6543 &19255 &19251 &19262\\
   \centering{\cite{Jin:2020jfc}}&6314&6375&6407&19237&19264&19279\\
   \centering{\cite{Berezhnoy:2011xn}} & 5966 &6051 & 6223&  18754 &18808 &18916\\
 \centering{\cite{Chen:2016jxd}}&6440–6470 &6370–6510 &6370–6510 &18460–18490 & 18320–18540&18320–18530\\  
  \centering{\cite{Karliner:2016zzc, Karliner:2020dta}} & $6192 \pm 25$ & $\cdots$ & $6429\pm 25$ & $18826 \pm 25$ & $\cdots$ & $18956 \pm 25$\\
%\cline{2-5}
 \centering{\cite{Wang:2018poa,Wang:2017jtz}}&5990$\pm$ 80 &6050$\pm$ 80 &6090$\pm$ 80 &18840 $\pm$ 90 & 18840$\pm$ 90 &18850$\pm$ 90\\ 
 \centering{\cite{Wu:2016vtq}}&7016 &6899 &6956 &20275 &20212 &20243\\ 
  \centering{\cite{Lundhammar:2020xvw}-Mod. I} & 5960 & 6009 & 6100 & 18723 & 18738 & 20243\\
%\cline{2-5}
\centering{\cite{Lundhammar:2020xvw}-Mod. II} & 6198 & 6246 & 6323 & 18754 & 18768& 18797\\
%\cline{2-5}
%\cline{2-5}
\centering{\cite{Zhao:2020nwy}} & 6476 & 6441& 6475 &19226& 19214& 19232 \\
  \centering{\cite{Bedolla:2019zwg}} & 5883 & 6120& 6246 &18748& 18828& 18900 \\ 
  \centering{\cite{Liu:2019zuc}} &6487 & 6500&  6524& 19322 &19329 &19341 \\
%\cline{2-5}
  \centering{\cite{Wang:2019rdo}}&6425&6425&6432&19247 &19247  &19249 \\
%\cline{2-5} 
  \centering{\cite{Deng:2020iqw}} &6407&6463&6486&19329&19373&19387\\
%\cline{2-5}
 \centering{\cite{Chen:2019dvd}}&$\cdots$ &$\cdots$ &$\cdots$ & 19178&19226&19236\\
  \centering{\cite{Lloyd:2003yc}}&6695 &6528 &6573 &$\cdots$ & $\cdots$&$\cdots$\\ 
\centering{\cite{Zhao:2020cfi}}&6480 &6508 &6565 &$\cdots$ & $\cdots$&$\cdots$\\ 
  \end{tabular}
\end{ruledtabular}
\end{table*}

Ref. \cite{Liu:2019zuc} used a nonrelativistic quark model within a potential model by including the linear confining potential, Coulomb potential, and spin-spin interactions in diquark-antidiquark picture. Ref. \cite{Wang:2019rdo} studied the mass spectra of the $S-$wave fully-heavy tetraquarks in two nonrelativistic quark models. Color-magnetic interaction model (CMIM), a traditional constituent quark model (CQM) and a multiquark color flux-tube model (MCFTM) were used to investigate the fully-heavy tetraquarks in diquark-antidiquark scenario \cite{Deng:2020iqw}. Ref. \cite{Chen:2019dvd} used nonrelativistic chiral quark model using the Gaussian expansion method for ground state of $bb\bar{b}\bar{b}$ tetraquark. Refs.  \cite{Wang:2017jtz,Wang:2018poa} made QCD Sum Rule studies on fully-heavy tetraquarks. In another work, only  fully-charmed tetraquarks were considered using a parameterized Hamiltonian with a sufficiently large but finite oscillator basis \cite{Lloyd:2003yc}. Ref. \cite{Zhao:2020cfi} calculated tetraquark masses in a constituent quark model, where the Cornell-like potential and one-gluon exchange spin-spin coupling were taken into account.

It can be seen from Tables \ref{tab:table9} and \ref{tab:table10} that  there are significant disagreements between the masses of fully-heavy $cc\bar{c}\bar{c}$ and $bb\bar{b}\bar{b}$ tetraquarks. Especially from Table \ref{tab:table9},  in the $cc\bar{c}\bar{c}$ sector, masses of $J^{PC}=0^{++}, 1^{+-}$, and $J^{PC}=2^{++}$ lie in the range of $5.8-7.0$ GeV, $6.0-6.8$ GeV and $6.1-6.9$ GeV, respectively. In the $bb\bar{b}\bar{b}$ sector, masses of $J^{PC}=0^{++}, 1^{+-}$, and $J^{PC}=2^{++}$   lie in the range of $18.7-20.0$ GeV, $18.3-20.2$ GeV and $18.3-20.2$ GeV, respectively. Refs. \citep{Debastiani:2017msn,Karliner:2020dta,Karliner:2016zzc,Wang:2017jtz,Wang:2018poa,Lundhammar:2020xvw,Bedolla:2019zwg} predict ground state fully-heavy  tetraquark masses below or slightly above the corresponding two meson threshold masses which means these states could be observed in the experiments. On the other side, present model of this work and other approaches predict ground state fully-heavy tetraquark masses significantly above the corresponding thresholds, and thus it is difficult to observe these states in the experiments. 2S fully-heavy tetraquark masses are significantly above then the corresponding lowest (1S-1S) two-meson threshold masses. Concerning excited 2S fully-heavy tetraquark masses, our results are significantly below than the corresponding two-meson thresholds and there can be an open window for observing these structures in the experiments. 

It would be good to analyze the response of the obtained results to minor variations of the numerical input of the potential model. The potential model has eight parameters. Therefore we divided into two group. In  the first group we have changed mass $m$ and $\alpha_s$ values and fixed the rest of parameters. In the second group we have changed coupling strength $h$, range $b$, $c$ and $V_0$ values and fixed the rest of parameters. The parameters of group 1 is shown in Table \ref{tab:table11}. 

\begin{table}[H]
\caption{\label{tab:table11} Group 1 potential parameters. }
\begin{ruledtabular}
\begin{tabular}{ccccc}
Input &Set 1&Set 2 & Set 3 & Set 4\\
\hline
$m_c$ & $1.41 ~ \text{GeV}$ & $1.39 ~ \text{GeV}$ & $1.4 ~ \text{GeV}$ &$1.4 ~ \text{GeV}$\\
$m_b$ & $4.822 ~ \text{GeV}$ & $4.802 ~ \text{GeV}$ & $4.812 ~ \text{GeV}$ &$4.812 ~ \text{GeV}$\\
$\alpha_s(m_c^2)$ & 0.37 & 0.37 & 0.38 & 0.36\\
$\alpha_s(m_b^2)$ & 0.26 & 0.26 & 0.27 & 0.25 \\
\end{tabular}
\end{ruledtabular}
\end{table}
As can be seen from Table \ref{tab:table11}, in Sets 1 and 2 we fixed $\alpha_s$ values constant and changed the mass values while in Sets 3 and 4, we fixed the mass values and changed $\alpha_s$ values. The results are given in Table \ref{tab:table12} for $cc\bar{c}\bar{c}$ tetraquark and in Table \ref{tab:table13} for $bb\bar{b}\bar{b}$ tetraquark.

\begin{table}[H]
\caption{\label{tab:table12} Responses in $cc\bar{c}\bar{c}$ tetraquark mass values to potential parameter input in group 1. PM stands for results of potential model used in this work. All results are in MeV.}
\begin{ruledtabular}
\begin{tabular}{ccccccc}
State&Set 1&Set 2 & Set 3 & Set 4& PM\\
\hline
$1^1S_0$ & 6331 & 6314 & 6320 & 6321& 6322\\
$1^3S_1$ & 6364 & 6342 & 6351 & 6353 &6354\\
$1^5S_2$ & 6393 & 6371 & 6383 & 6384& 6385\\
\end{tabular}
\end{ruledtabular}
\end{table}

\begin{table}[H]
\caption{\label{tab:table13} Responses in $bb\bar{b}\bar{b}$ tetraquark mass values to potential parameter input in group 1. PM stands for results of potential model used in this work. All results are in MeV.}
\begin{ruledtabular}
\begin{tabular}{ccccccc}
State&Set 1&Set 2 & Set 3 & Set 4 & PM\\
\hline
$1^1S_0$ & 19687 & 19646 & 19665 & 19666 & 19666\\
$1^3S_1$ & 19698 & 19654 & 19672 & 19673 & 19673 \\
$1^5S_2$ & 19708 & 19759 & 19680 & 19680 & 19680\\
\end{tabular}
\end{ruledtabular}
\end{table}
As can be seen from Tables \ref{tab:table12} and \ref{tab:table13}, the responses in the numerical input changes of group 1 did not significantly effect the results. The parameters of group 2 are given in Table \ref{tab:table14}. 

\begin{table}[H]
\caption{\label{tab:table14} Group 2 potential parameters. }
\begin{ruledtabular}
\begin{tabular}{ccccc}
Input &Set 5&Set 6 & Set 7 & Set 8\\
\hline
$h$&$0.21 ~ \text{GeV}$ & $0.19 ~ \text{GeV}$ & $0.20 ~ \text{GeV}$  &$0.20 ~ \text{GeV}$ \\
$b$  & $0.5 ~ \text{GeV}$ & $0.3 ~ \text{GeV}$ & $0.4 ~ \text{GeV}$ &$0.4 ~ \text{GeV}$\\
$c$  & $0.193 ~ \text{GeV}^2$ & $0.193 ~ \text{GeV}^2$ & $0.203 ~ \text{GeV}^2$ & $0.183 ~ \text{GeV}^2$\\
$V_0$ & $-0.223 ~ \text{GeV}$  & $-0.223 ~ \text{GeV}$  & $-0.233 ~ \text{GeV}$  & $0.213 ~ \text{GeV}$\\
\end{tabular}
\end{ruledtabular}
\end{table}
As can be seen from Table \ref{tab:table14}, in Sets 5 and 6 we fixed $c$ and $V_0$ values constant and changed $h$ and $b$ values while in Sets 7 and 8, we fixed  $h$ and $b$ values and changed $c$ and $V_0$ values. The results are given in Table \ref{tab:table15} for $cc\bar{c}\bar{c}$ tetraquark and in Table \ref{tab:table16} for $bb\bar{b}\bar{b}$ tetraquark. 

\begin{table}[H]
\caption{\label{tab:table15} Responses in $cc\bar{c}\bar{c}$ tetraquark mass values to potential parameter input in group 2. PM stands for results of potential model used in this work. All results are in MeV.}
\begin{ruledtabular}
\begin{tabular}{ccccccc}
State&Set 1&Set 2 & Set 3 & Set 4 & PM\\
\hline
$1^1S_0$ & 6314 & 6329 & 6330 & 6314 & 6322 \\
$1^3S_1$ & 6345 & 6372 & 6362 & 6344  &6354\\
$1^5S_2$ & 6376 & 6401 & 6394 & 6375 & 6385\\
\end{tabular}
\end{ruledtabular}
\end{table}

\begin{table}[H]
\caption{\label{tab:table16} Responses in $bb\bar{b}\bar{b}$ tetraquark mass values to potential parameter input in group 2. PM stands for results of potential model used in this work. All results are in MeV.}
\begin{ruledtabular}
\begin{tabular}{ccccccc}
State&Set 1&Set 2 & Set 3 & Set 4 & PM\\
\hline
$1^1S_0$ & 19676 & 19680 & 19667 & 19664 & 19666\\
$1^3S_1$ & 19682 & 19686 & 19674 & 19672 & 19673 \\
$1^5S_2$ & 19689 & 19694 & 19680 & 19768 & 19680\\
\end{tabular}
\end{ruledtabular}
\end{table}
As can be seen from Tables \ref{tab:table15} and \ref{tab:table16}, the responses in the numerical input changes of group 2 did not significantly effect the results.

We believe that, the results of this work should serve as a benchmark for other studies, which are necessary to identify and develop theoretical models. This effort is important since predictions with respect to different models or approaches should be in well agreement and depend mildly to the corresponding model parameters.

\section{\label{sec:level4}Summary and Concluding Remarks}
In this work, adapting a perspective in which tetraquarks are assumed to be compact and consist of axial-vector diquarks and antidiquarks, masses of fully-heavy tetraquark states were calculated. We used a nonrelativistic potential quark model with the linear confining potential, Hulthen potential, and spin-spin interactions. We predicted ground, first and second radially excited mass spectra of fully-heavy tetraquarks, $cc\bar{c}\bar{c}$ and $bb\bar{b}\bar{b}$. In order to do this, we adopted a model which was formulated for quarkonia systems and made use of this model for diquark and tetraquark systems. 

We have factorized four-body system into two two-body systems: quarks form diquark and antiquarks form antidiquarks, then diquark and antidiquark form tetraquark. By doing this, the complex structure of four-body problem became more accessible. This flowchart  allows us to study both ground and excited tetraquark states. Furthermore, various spin configurations were conducted easily in this manner.

At first step, we obtained diquark/antidiquark masses. Since diquark masses equal antidiquark masses, in the second step we computed tetraquark masses. In comparison with other models, the obtained $cc$ and $bb$ diquark  masses are in good agreement with  earlier results. This situation can be seen in Tables \ref{tab:table5} and \ref{tab:table7}. Indeed, radial excited states of $cc$ and $bb$ diquark  masses agree well with available results in the literature which can be seen in Tables \ref{tab:table4} and \ref{tab:table6}. After obtaining diquark masses, we computed ground, first, and second radially excited $cc\bar{c}\bar{c}$ and $bb\bar{b}\bar{b}$ tetraquark masses. The results are presented in Table \ref{tab:table8}. It was found that both ground states of $cc\bar{c}\bar{c}$ and $bb\bar{b}\bar{b}$ tetraquarks masses are considerably higher than their corresponding lowest (1S-1S) two-meson decay threshold masses. Therefore they might easily decay into the lowest two-meson allowed states and these decays proceed through quark rearrangements. Such decays are not suppressed dynamically or kinematically. These states should be broad and are difficult to observe experimentally. We observed that first excited states of fully-charmed and fully-bottomed tetraquarks are lower than their corresponding (2S-2S) two-meson thresholds. These tetraquarks might be observed  by obtaining enough center of mass energy and luminosity in the experiments. We also observed that the masses of 2S  $cc\bar{c}\bar{c}$ and $bb\bar{b}\bar{b}$ tetraquarks are significantly higher than the corresponding lowest (1S-1S) two-meson threshold masses. First and second radially excited masses of fully-heavy tetraquarks are compatible with the reference studies given in Table \ref{tab:table9}.

The comparison of ground state masses of $cc\bar{c}\bar{c}$ and $bb\bar{b}\bar{b}$ tetraquarks are presented in Table \ref{tab:table10}. As argued before, the results deviate up to 1.0 GeV for $cc\bar{c}\bar{c}$ tetraquark and 2.0 GeV for $bb\bar{b}\bar{b}$ tetraquark which are considered as high differences. Significant deviations between diquark and tetraquark masses may present a deeper perspective. In quark model, quarks are point-like objects.  The dynamics, interactions and bound states for two and three quarks systems are described well. The diquark-antidiquark models require additional assumptions. In this perspective most diquark-antidiquark models assume diquarks and antidiquarks are point-like objects so that the interactions are effectively the same as for ordinary quarkonia systems. This was argued, for example in Ref. \cite{Faustov:2020qfm}, in which diquarks have a structure (size) and this effect the results. In addition to this, chromomagnetic interaction inside the diquarks or between diquarks are hypothesized \cite{Maiani:2014aja,Richard:2017vry}.

To sum up, we believe that our predictions 	for fully-charmed and fully-bottomed ground and excited states may help to experimental studies for a possible tetraquark signal in corresponding channels.

\bibliography{apssamp}

\end{document}